\documentclass[prl,twocolumn,showpacs,preprintnumbers]{revtex4}
\usepackage{graphicx} 
\usepackage{dcolumn}
\usepackage{bm}
\usepackage{booktabs} 
\usepackage{amsmath}  
\usepackage[colorlinks=true,linkcolor=red]{hyperref} 
\usepackage{memhfixc} 

\begin{document}

\title{Evolution in Surface Morphology of Epitaxial Graphene Layers on SiC Induced by Controlled Structural Strain}
\author{Nicola Ferralis}
\email{nferralis@berkeley.edu}
\affiliation{Department of Chemical Engineering, University of California,
 Berkeley, California 94720}
\author{Jason Kawasaki}
\altaffiliation[Current address: ]{Department of Mechanical and Aerospace Engineering, Princeton University, Princeton, NJ}
\affiliation{Department of Chemical Engineering, University of California,
 Berkeley, California 94720} 
\author{Roya Maboudian}
\affiliation{Department of Chemical Engineering, University of California,
 Berkeley, California 94720}
\author{Carlo Carraro}
\affiliation{Department of Chemical Engineering, University of California,
 Berkeley, California 94720}

\date{\today}

\begin{abstract}
The evolution in the surface morphology of epitaxial graphene films and  6H-SiC(0001) substrates is studied by electron channeling contrast imaging. Whereas film thickness is determined by growth temperature only, increasing growth times at constant temperature affect both internal stress and film morphology. Annealing times in excess of 8-10 minutes lead to an increase in the mean square roughness of SiC step edges to which graphene films are pinned, resulting in compressively stressed films at room temperature. Shorter annealing times produce minimal changes in the morphology of the terrace edges and result in nearly stress-free films upon cooling to room temperature.

\end{abstract}

\pacs{81.05.Uw, 65.40.De, 68.37.Hk, 78.67.-n, 78.30.-j}

\maketitle

The remarkable electronic properties of graphene have shown promise for use in high performance electronic devices \cite{Novoselov_2004}. However, large area fabrication processes for graphene devices have not yet been developed. Epitaxial graphene, which shares key electronic properties of free-standing graphene, has emerged as an attractive alternative to the layer-by-layer exfoliation process \cite{Berger_2004, Berger_2006, Hass_2006, Ohta_2006, deHeer_2007, Hass_2008}, because of its potential for scalability and for integration with Si-based electronics. Epitaxial graphene layers were shown recently to display blueshifts in the frequencies of their optical phonons at room temperature \cite{Ferralis_2008, Rohrl_2008}, which are indicative of the presence of lattice strain in the epilayer. This strain is explained by the large difference in the thermal expansion coefficients of graphene and SiC which is nearly constant from the synthesis to room temperature \cite{Ferralis_2008, Rohrl_2008}. 

Strained heterostructures have received a significant amount of interest in the past because of the wide ranging implications of strain, from the ability to engineer bandgaps in silicon-based materials for electronic device fabrication \cite{Maiti_2004} to the constraints it poses on the growth of defect-free heterostructrures \cite{Jain_1994}. It is therefore of interest to investigate some of the consequences of the presence of strain on epitaxial graphene film properties and on epilayer quality. It is known that the amount of residual strain at room temperature in epitaxial graphene is tunable by adjusting the annealing time \cite{Ferralis_2008}. In this paper we investigate the evolution in the morphology of graphene domains pinned at SiC terrace edges with increased annealing time. An increase in the step edge mean square roughness is observed with annealing time. The evolution in step morphology is proposed as a mechanism for the relief of structural strain in the graphene epilayer, which is, in turn, controlled by the annealing time. 

Graphene epilayers were produced by thermal annealing the Si-face of 6H-SiC single crystals with (0001) orientation (CREE Research, Inc.). 6H-SiC samples were first prepared {\it ex vacuo} by high temperature annealings in hydrogen athmosphere to obtain a regularly stepped surface, as described elsewhere \cite{Ferralis_2008}. The stepped surface is necessary to monitor the morphological evolution of epitaxial graphene as it grows off SiC terrace edges \cite{Hannon_2008}. The resulting SiC surface was characterized by a sharp $(3\times 3)$ low-energy electron diffraction (LEED) pattern. Graphene layers were grown by annealing the SiC substrate in ultrahigh vacuum (UHV) for varied times (ranging from 1 to 30 minutes); subsequently, the substrate was cooled down to room temperature at the rate of $\le$ 1 Ks${}^{-1}$. The same growth temperature $T_{G}= 1250\pm20^{\circ}$C was used, in order to obtain the same number of graphene epilayers \cite{Berger_2004, deHeer_2007}. The film thickness was measured after the annealing by monitoring the C:Si Auger electron spectroscopy (AES) ratio \cite{deHeer_2007}. In all data presented here on the graphitized substrate, the C:Si ratio was 7$\pm$1, corresponding to 1.0$\pm$0.2~monolayer~(ML). At room temperature, the resulting graphene layers possess a threefold symmetric LEED pattern superimposed on the (6$\sqrt 3\times 6\sqrt 3) R30^{\circ}$ pattern of the SiC surface reconstruction \cite{Forbeaux_1998, Berger_2004}. Surface morphology was monitored with electron channeling contrast imaging (ECCI), through a low-voltage scanning electron microscope (NovelX MySEM), operating at 1kV in topographic mode. The ability to directly image atomic terrace edges with ECCI has been demonstrated recently for crystalline SiC surfaces \cite{Picard_2007}.

\begin{figure*}
\includegraphics{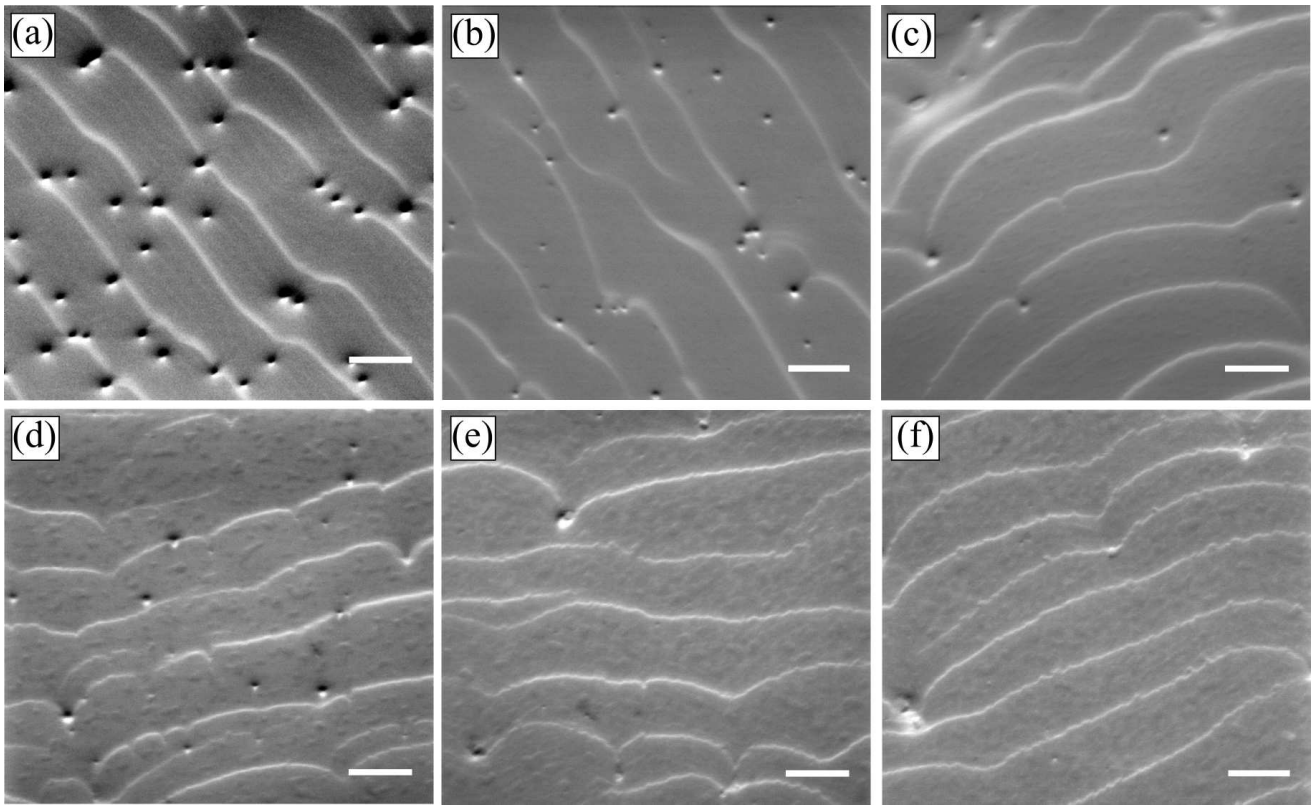}
\caption{\label{Fig1}Electron channeling contrast images of 1~ML epitaxial graphene layers synthesized at different annealing times: (a) clean SiC surface (0 min); (b) 1 min; (c) 2 min; (d) 4 min; (e) 8 min; (f) 30 min. Scale bar: 1~$\mu$m.}

\end{figure*}

\begin{figure}
\includegraphics{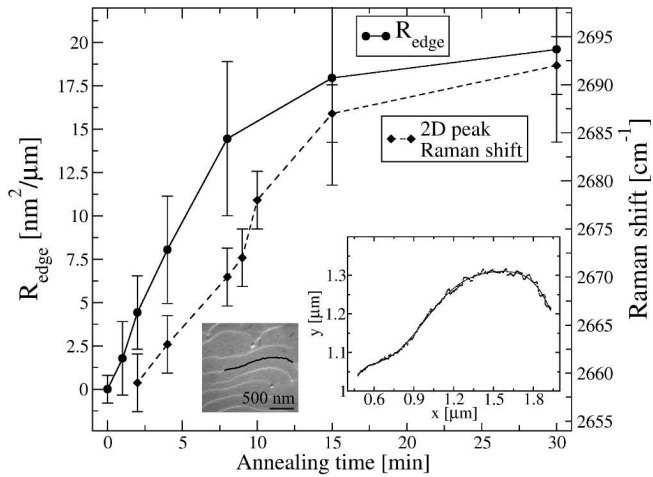}
\caption{\label{Fig2} The edge roughness R$_{edge}$ (defined as the difference in the normalized average mean square deviation of any graphitized terrace edge with that of the initial ungraphitized surface) and the 2D Raman shift line \cite{Ferralis_2008} are shown as function of annealing time. Several profiles of terrace edges are extracted from SEM images of samples prepared with the same annealing time and temperature. Each profile (black curve on the SEM image of a sample annealed for 8 minutes) is fit with a 9$^{th}$ order polinomial to obtain an edge baseline. The normalized average mean square deviation (and thus the edge roughness R$_{edge}$) is extracted from the baseline.}

\end{figure}

Figure~\ref{Fig1}(a) shows ECCI images of the SiC surface after the high temperature annealing in hydrogen ambient, but before the high temperature graphitization in  (UHV). Two sets of terrace edges are visible: major step bunches spaced about 1.1~$\mu$m, and subterraces spaced about 200~nm. The major terraces are 0.8~nm thick (measured by atomic force microscopy), corresponding to 3 SiC bilayer, in agreement with previous studies \cite{Hannon_2008}. Terrace edges of the step bunches in the clean surface appear smooth and the subterraces can be clearly resolved. Scanning electron micrographs of 1~ML graphene grown in UHV for different annealing times at $T_{G}= 1250^{\circ}$C are shown in Figs.~\ref{Fig1}(b)-(e). With increased annealing time, the major terrace edges become visibly rougher, while the subterraces are much less clearly resolved. The edge roughening becomes increasingly more evident for longer annealing times. To quantify this effect, the mean square roughness of the step bunches as a function of annealing time was measured. Several terrace edges of an individual sample were traced using plot digitizing software at a sampling rate of about 17~nm/step (inset in Fig.~\ref{Fig2}). The trace was then fit against a 9$^{th}$ order polynomial, which provides the baseline fit $y_0(t)$. The resulting baseline was then used to extract the average mean square (MS) deviation from the baseline $y_0(t)$. The average MS deviation was found to be proportional to the edge length $L$. Thus, for each terrace edge, the average MS deviation was normalized by $L$. Finally, the edge roughness R$_{edge}$ was defined as the difference in the normalized MS deviation of any graphitized terrace edge with that of the initial ungraphitized surface (so that R$_{edge}$ for the initial surface is zero). The evolution in R$_{edge}$ with annealing time is shown in Fig.~\ref{Fig2}.

A sharp increase in step edge roughness occurs during the first 8-10 minutes of annealing, when R$_{edge}$ rises to about 15~$nm^2/\mu m$. For longer annealing times the edge roughness increases more gradually to a maximum of about 20~$nm^2/\mu m$. In addition to edge roughening, the SiC subterraces become increasingly rougher and finally appear to aggregate into large patches (Fig.~\ref{Fig1}(d)-(f)). It has been shown recently that graphene layers are formed by decomposition of the $\sqrt3$ terraces \cite{Hannon_2008}. This process develops until much of the higher SiC step edge is consumed to form a graphene layer on the lower terrace. It is very likely that the variation in the terrace edge morphology oberved here is related to the local dissolution of the SiC steps and the formation of graphene layers. However, such variation at constant growth temperature, does not appear to alter the surface coverage considerably ($\pm$0.2 ML, based on AES).

With increased annealing times, a similar trend of edge roughness is observed in the evolution in the 2D Raman line (Fig.~\ref{Fig2}), which indicates an increase in the strain of the graphene epilayer {\it measured at room temperature}.
Uniform compressive strain in epitaxial graphene originates in the large difference between the coefficients of linear thermal expansion of SiC and graphene \cite{Ferralis_2008}, and in the fact that graphene expands upon cooling \cite{Mounet_2005}, while SiC contracts. The difference $\Delta\alpha(T)$ in the coefficients of thermal expansion is nearly constant between room temperature (RT) and the graphene synthesis temperature, $T_{G}\approx 1250^{\circ}$C. A higher compressive stress  at room temperature results from a lower stress film at the temperature of growth ($T_{G}$), while a nearly stress free film at room temperature indicates that the film existed under high tensile stress at $T_{G}$. Within experimental accuracy, the strain measured at room temperature might well vanish for very short annealing times. In contrast, for long annealing times, assuming that the graphene layer is in mechanical equilibrium with the substrate at the temperature of growth, a compressive strain in the room temperature film of ~0.8\% is calculated, which is in agreement with recent measurements \cite{Ferralis_2008, Rohrl_2008}. This analysis suggests that mechanical equilibrium with the 6-$\sqrt3$ SiC substrate at $T_{G}$ is indeed achieved for annealing times longer than 10 minutes, while for shorter annealing times ($\sim$5 minutes or less), graphene is under high tensile strain at $T_{G}$. 

The strain relaxation at $T_{G}$, implied by the Raman data, provides a likely explanation for the changes in morphology observed by ECCI. Accordingly, for long enough annealing time, tensile strain developed at $T_{G}$ is relieved by the break up of the pinned graphene layer on the SiC subterraces  into a patchy film, as observed. These changes in morphology must be accompanied by roughening of the step edges to which graphene films are pinned. Such increase in roughness does not induce a significant change in surface coverage ($\pm$0.2~ML). For shorter annealing times, surface relaxation does not take place, leaving the SiC terraces morphologically unchanged. 

In summary, the evolution in the surface morphology of epitaxial graphene films and of the underlying (0001) 6H-SiC substrate was studied by ECCI and explained in terms of the changes in structural strain during the synthesis of epitaxial graphene. It is found that long annealing times produce layers in equilibrium at growth temperature (and thus under highly compressive strain at room temperature); stress relief at the growth temperature is obtained through the development of large step roughness (as high as 20~$nm^2/\mu m$) and film discontinuity. Shorter annealing times produce highly tensile films at growth temperature, resulting in almost stress-free films at room temperature as well as minimal change in surface morphology. Thus the ability to control strain in epitaxial graphene, by tuning of the annealing times, is tied with the ability to control the film morphology.

\begin{acknowledgments}
Support of the National Science Foundation (Grants EEC-0425914 and CMMI-0825531), and the DARPA-MTO is gratefully acknowledged.
\end{acknowledgments}

\end{document}